\catcode`\@=12
{\count255=\time\divide\count255 by 60 \xdef\hourmin{\number\count255}
        \multiply\count255 by-60\advance\count255 by\time
   \xdef\hourmin{\hourmin:\ifnum\count255<10 0\fi\the\count255}}
\def\ps@draft{\let\@mkboth\@gobbletwo
    \def\@oddhead{}
    \def\@oddfoot
       {\hbox to 7 cm{$\scriptstyle Draft\ version:\ \draftdate$
       \hfil}
       \hskip -7cm\hfil\rm\thepage \hfil}
    \def\@evenhead{}\let\@evenfoot\@oddfoot}
\catcode`\@=12
\def\draftdate{\number\month/\number\day/\number\year\ \ \ \hourmin }
\def\draft{\pagestyle{draft}\thispagestyle{draft}}
\documentstyle[12pt]{article}
\newcommand{\be}{\begin{eqnarray}}
\newcommand{\en}{\end{eqnarray}}
\newcommand{\non}{\nonumber}
\newcommand{\no}{\noindent}
\newcommand{\vs}{\vspace}

\newcommand{\p}{\partial}
\newcommand{\ha}{{1\over 2}}

\title{{\bf Renormalization Group and Asymptotics of Solutions
of \\ Nonlinear Parabolic Equations}}

\author{J.Bricmont\\UCL, Physique Th\'eorique,
 B-1348, Louvain-la-Neuve, Belgium\and
A.Kupiainen\thanks{Supported by  NSF grant DMS-8903041}
\\Rutgers University, Department of Mathematics ,\\ New Brunswick
NJ 08903, USA\and
G.Lin\\ Rutgers University, Department of Mathematics ,\\ New Brunswick
NJ 08903, USA}

\date{}

\begin{document}

\maketitle
\begin{abstract}
We present a general method for studying long time asymptotics
of nonlinear parabolic partial differential equations. The
method does not rely on a priori estimates such as the maximum
principle. It applies to systems of coupled equations,
to boundary conditions
at infinity creating a front, and to higher
(possibly fractional) differential
linear terms. We present in detail the analysis for
nonlinear diffusion-type equations with initial data falling
off at infinity and also for data interpolating between two
different stationary solutions at infinity.
In an accompanying paper \cite{BK},
the   method is applied to systems of equations where
some variables are "slaved", such as the complex Ginzburg-Landau
equation.
\end{abstract}

 \no  {\Large\bf 1. Introduction  }

\vs{0.5 cm}
\draft

The time evolution of many physical quantities is described
 by nonlinear, parabolic, partial
differential equations. For most of these equations, to obtain
 a closed form solution seems
to be a
hopeless task. Therefore, one tries to determine certain
qualitative properties of the
solution, such
 as its existence and regularity for all times, or its long-time
asymptotics. It turns out that,
for  certain equations, the long-time behaviour can be
predicted because the solution becomes
asymptotically scale-invariant; a trivial example is given
by the usual
heat equation: $\dot{u}=u''$. The fundamental solution
$u(x,t)= (4\pi t)^{-{1\over 2}}e^{-{x^2\over
4t}}$ is called scale invariant because its value at time
$t'$ may be obtained from its value at
time $t$ by scaling $x$ and $u$. Moreover, for (suitable)
initial data $u_0 (x)$, the solution is
asymptotically proportional to the fundamental
solution, and therefore decays  like
$ t^{-{1\over 2}}$. A less trivial remark is that
the same asymptotic
behaviour governs other equations; consider, for example,
the heat equation
with absorption: $\dot{u}=u'' - u^{p}$. Then, for $p>3$,
the long time
asymptotics, given integrable initial data, is again given by the
fundamental solution of
 the heat
equation \cite{Ga,GV,KP1}. For long times, the nonlinear term $u^{p}$
has   almost no effect.

A similar situation is encountered
in statistical mechanics in
the theory of critical phenomena. Here too, it seems
 difficult to obtain an exact solution for
many systems of interest. However, at the critical point,
correlation functions become
asymptotically scale-invariant; moreover, universality holds
i.e. one may partition  the set of
statistical systems into different classes in such a way that,
  within a class, different systems
have the same asymptotic behaviour at the critical point.

	In the theory of critical phenomena, scale invariance and
  universality can both be understood
on the basis of renormalization group transformations. The idea is,
very
roughly, as follows: start  from, say, a lattice system
(like the Ising model),
defined by its Hamiltonian $H$, and fix a parameter
 $L$. Then, perform the statistical sum over all degrees of
freedom corresponding to fluctuations
of  scale less than $L$. Rescale everything by $L^{-1}$; we obtain
a new lattice system defined by a
renormalised
 Hamiltonian $H'=R_{L} (H)$. This defines the renormalization group
 transformation.
Now, iterate this procedure. The action of this transformation on
the space of
Hamiltonians is expected to have the following structure (this can
 be proven in
certain cases): if we start with
 a Hamiltonian $H$ which correspond to a critical point, the
renormalization
group will drive it  towards a fixed point of that transformation. The
fixed
point is scale-invariant, which accounts for the asymptotic scale
invariance of
the original system, and a universality class corresponds
 to the basin of attraction of a given fixed point. Moreover,
it often suffices to do a linear
stability analysis of the RG transformation around the fixed
 point in order to determine
qualitatively  its basin of attraction.

 From a more technical point of view, the renormalization group
allows the use of
perturbation  theory even where it seems to fail: in spin systems,
there exist high and low
temperature  expansions which give good results away from the
critical point but fail to
converge up  to that point. However such perturbative methods
can be used to study the RG map
itself,  because the latter involves only the summation over
      a finite number of degrees of
freedom  (those of scale less than $L$).

 In 1978, Barenblatt wrote a book \cite{Ba} devoted to the
role of scaling (which he calls ''self-similarity of the second
kind") in the asymptotic behaviour of solutions of PDE's. He
suggested to look
at these equations using the ideas being developed in field theory
and in the theory of phase transitions. However, it is only recently,
that, in
several very interesting papers \cite{G1,G2,G3}, it was shown how
the   RG idea can be used to study the long-time  asymptotics of
parabolic PDE's.  The purpose of this paper is to extend   this
analysis and  to show that it  yields new mathematical results. For
example, we can show
 that, for a  large class of
perturbations of the heat equation and for small
 initial data,  the long time
behaviour of the solution is given by the fundamental
solution of the heat  equation. This
result, which is related to those of \cite{Ga,GV,KP1,CEE},
also holds if we replace
the second derivative   in the heat equation  by
some other derivatives.

We can also study what happens when one takes initial data
 that have a non trivial behaviour
at infinity, i.e. such that $ u_0 (x)$
  tends to two different stationary solutions of
the PDE when $x$ tends to infinity .  We show again in a
general situation, with small initial data,
 that the solution acquires in the
long time limit a universal form, a universal "front",
whose shape depends only on the boundary conditions and
on the "universality class" of the equation.
In an accompanying
paper \cite{BK} we extend this result to the
"slaving principle", i.e. to the fact that some (fast) variables
 in coupled systems are essentially following the
behaviour of other (slow) variables. Those fast variables
are an example of what is called relevant variables in
RG terminology. As an example we consider in \cite{BK}
the complex Ginzburg-Landau equation (see also \cite{CE2,CEE}).

The RG transformation for the PDE's is simply
the integration
of the equation up to an
$L$-dependent time, followed by rescaling.
After rescaling, we get
 a new equation. Upon iteration, this tends
 to stabilise to a
fixed equation,
whose scale invariant solutions are the fixed points
of the RG transformation.
Here also, the power of perturbation theory is
enhanced by the RG approach:
while the  standard problem is to justify
 perturbation theory for an
infinite
time, here we use it  to study the RG map,
which is only a finite time
problem.
However, as in the theory of  critical
 phenomena, the RG idea is not
limited in
principle to the perturbative regime.

We hope that the RG method will be a fruitful
 way to analyse a general class of
equations.
In any case, the equations discussed below provide very simple
illustrations of rigorous applications of
the RG method. We have tried to
stress the analogy between the partial
 differential equations analysed here
and
technically more complicated applications of the RG
like field theory (see Section 5).

\vspace{1cm}

\no  {\Large\bf 2. The Renormalization Group Map  }

\vs{0.5 cm}

We will explain now the RG method for an equation of the type
\be
\dot{u}=u''+F(u,u',u'')
\en
(for notational simplicity we take the spatial variable $x\in
{\bf R}$: everything below can as well be done in ${\bf R}^n$).
We are interested in the asymptotics of the solution of the form
\be
u(x,t)\sim t^{-{\alpha\over 2}}f(xt^{-{1\over 2}})
\en
as $t\rightarrow\infty$.

The standard way to study this problem is, \cite{Ga,GV,KP1}, to first
look for a scale-invariant solution (if F is scale-invariant)
which reduces to solving an ordinary differential equation (see
Section 5) and then to establish the
stability of this solution. Usually, one uses
 a priori estimates like
the maximum principle to prove stability.

The RG method transforms the problem of large
time limit into an
iteration of a fixed time problem  followed by
a scaling transformation. The scale-invariant solution emerges
then as a fixed point of a map in the space of initial data,
the RG map, and the stability analysis becomes the
analysis of the stability of the fixed point under the RG.
As we will see, the method does not depend on
any a priori positivity
properties and is applicable to systems of
equations
and to equations whose leading term has other
derivatives.

We now explain the idea without
spelling out any concrete assumptions on the
$F$ in (1) nor on the spaces and norms. The choice of
the latter, as usual in
non linear problems, depends on the particular problem.
Thus, let us fix some (Banach) space of initial data $\cal S$.
It will be convenient to take the $initial$ $time$ to be $t=1$.
Next,
we pick a number $L>1$ and set
\be
u_L(x,t)=L^\alpha u(Lx,L^2t)
\en
where $\alpha$ will be chosen later and $u$ solves (1) with the
initial data $f\in\cal S$. The RG map $R:{\cal S}\rightarrow
{\cal S}$ (this has to be proven!) is then
\be
(Rf)(x)=u_L(x,1) .
\en
Note that $u_L$ satisfies
\be
\dot{u}_L=u''_L+F_L(u_L,u'_L,u''_L) .
\en
with $F_L(a,b,c)=L^{2+\alpha}F(L^{-\alpha}a,L^{-1-\alpha}b,
L^{-2-\alpha}c)$.

We may now iterate $R$ to study the asymptotics of (1).
$R$ depends, besides on $\alpha$, on $L$ and $F$. Let us denote this
by $R_{L,F}$. We have then the "semigroup property"
\be
R_{L^n,F}=R_{L,F_{L^{n-1}}}\circ\dots\circ R_{L,F_L}\circ R_{L,F}.
\en
Each $R$ on the RHS involves a solution of a fixed time problem and
the long time problem on the LHS is reduced to an iteration of these.
Letting $t=L^{2n}$, we have
\be
u(x,t)=t^{-{\alpha\over 2}}(R_{L^n,F}f)(x t^{-{1\over 2}}).
\en

Now one tries to show that there exists an $\alpha$ such that
\be
F_{L^{n}}\rightarrow F^{\ast}\; , \hspace{3mm}
R_{L^n,F}f\rightarrow f^{\ast}
\en
where
\be
R_{L,F^{\ast}}f^\ast =f^\ast
\en
is the fixed point of the RG, corresponding
to the scale-invariant
equation $\dot{u}=u''+F^\ast$. Then, rescaling $x$,
the asymptotics of the original
problem is given by
\be
u(x t^{1/2},t)\sim t^{-{\alpha\over 2}}f^\ast (x).
\en
We will now illustrate the RG method in two concrete cases.

\vspace{1cm}

\no {\Large\bf 3. Gaussian Fixed Point: Diffusive Repair}

\addtocounter{equation}{-10}

\vspace{5mm}

For $F=0$ the equation (2.1) is just the diffusion equation and
the corresponding fixed points are trivial to write down. We
concentrate on integrable initial data; for a discussion of
fixed points relevant to other data see the end of this section.

Going to Fourier transform and putting $\alpha=1$, we have
\be
\widehat{Rf}(k)=e^{-k^2(1-L^{-2})}\hat{f}(L^{-1}k)\equiv
\widehat{R_0 f}(k).
\en
$R_0$ has a line of fixed points, namely the multiples of $f^*_0$, with
\be
\hat f^*_0(k)=e^{-k^2}.
\en
$f^*_0$ is of course the initial data of the scale
invariant solution $u(x,t)= (4\pi t)^{-{1\over 2}}e^{-{x^2\over
4t}}$.

Note that $R_0$ has nice contractive properties, provided
$\hat{f}$ has some smoothness. Indeed,
let $f=f^*_0+g$ with $\hat{g}(0)=0$. Then, $R_0 f=
f^*_0 + R_0 g$ and, if $\hat{g}$ is $C^1$, we have
$|\hat{g}({k\over L})|\leq {|k|\over L}|\hat{g}'
({ks\over L})| $, for $0 \leq s \leq 1$,
which leads to contraction in many "natural" norms.
 Here, we shall consider
the Banach space ${\cal B}$ of functions $f$ with
                             $\hat{f}\in C^1({\bf
R})$, equiped with the norm (we are inspired here by \cite{CEE})
\be
\|f\|=\sup_k (1+k^4)(|\hat{f}(k)|+ |\hat{f}'(k)|).
\en
The $1+k^4$ factor provides smoothness for $f$ (actually we
only need $1 + |k|^p$ with $p>3$) whereas the
$\hat{f}'$ term gives
decay at infinity (e.g. $xf(x)\in L^2$). Now, for $\hat{g}(0)=0$
as above, we get
\be
\|R_0 g\|\leq  CL^{-1}\|g\|,
\en
with $C$ independent of $L$. Here and below, $C$ will denote a
generic constant, which may change from place to place, even in
the same equation. However, we shall write $C_L,C_{F}$,... for (generic)
constants that depend on the choice of $L$, $F$ etc. In the proofs,
$L$ will be chosen large enough so that e.g. $CL^{-1} <1$. Then other
quantities like $\epsilon$ in Theorem 1 below are chosen small enough
in relation with $L$.

Next, we discuss the domain of attraction of the Gaussian fixed
point (2) in the set of equations of the form (2.1).
We will take the nonlinear
term $F$ to be a function $F:{\bf C}^3\rightarrow {\bf C}$ which is
analytic in a neighbourhood of the origin. Note that for a monomial
$F(a,b,c)=a^{n_1} b^{n_2} c^{n_3}$ we have
$F_L=L^{-d_F}F$ with
\be
d_F=n_1+2 n_2+3 n_3 -3.
\en
In general, we define $d_F$ for an analytic $F$ by taking the
smallest of the numbers (5) computed for the monomials in the
Taylor series of $F$ at 0 with non-vanishing coefficients.
We call $F$ $irrelevant$, if $d_F>0$, $marginal$, if $d_F=0$
and $relevant$, if $d_F<0$.

\vspace{2mm}

\no {\bf A. The irrelevant case}

\vspace{2mm}

We can now state the main result for an irrelevant $F$.

\vspace{2mm}

\no {\bf Theorem 1}. {\it Let $F:{\bf C}^3\rightarrow
{\bf C}$ be analytic
in a neighbourhood of $0$ with $d_F>0$ and fix a $\delta>0$.
Then there exists an $\epsilon>0$
such that if $\|f\|<\epsilon$, the equation
\be
\dot{u}=u''+F(u,u',u'') {\hspace{3mm}}u(x,1)=f(x)
\en
has a unique solution which satisfies, for some number $A =A(f,F)$,}
\be
\lim_{t\rightarrow\infty} t^{1-\delta}\|u(\cdot
t^{1\over 2},t)-At^{-{1\over 2}}
f^*_0(\cdot )\|=0.
\en

\vspace{2mm}

\no{\bf Remark 1}. As will be evident from the proof, we could
as well consider a more general class of equations
\be
\dot u = -(-\Delta)^{\beta / 2} u + F \; ,\;\;\;
u:{\bf {R}}^N\rightarrow {\bf C}
\en
where $\widehat{(-\Delta)^{\beta/2} f}(k)\equiv
|k|^\beta\hat{f}(k)$
and $F$ is analytic in $u$ and its
partial derivatives
up  to order $\beta$. $\hat{f}^*(k)$ is then $e^{-|k|^\beta}$ and
$u(x,t)\sim At^{-{N\over\beta}}f^*(x t^{-{1\over\beta}})$,
provided $d_F >0$ where, for a monomial $F=\prod^N_{j=1}
\prod^l_{i=1} (\p_j^{a_j}u)^{n_{ij}}$, we set
\be
d_F = \sum^N_{j=1} \sum^l_{i=1} (N + a_i ) n_{ij} - (N+
\beta)\non.
\en

We restrict $a_i \leq \beta$. In (2.3) we set $\alpha=N$ and
replace $L^2 t$ by $L^\beta t$. In (3), we replace $k^4$
by $| k |^p$ with
$p > \beta + N$. In (5) above, we had $N=1,\beta=2$
and $a_i=0,1,2$. Of course we could also take in $F$ fractional
derivatives, $|k|^\gamma$ in momentum space for $\gamma\leq\beta$.

\vspace{2mm}

\no {\bf Remark 2.} The statement (7) translates in momentum
space into

\be
|\hat{u}(k,t)-Ae^{-tk^2}|\leq Ct^{-\ha +\delta}(1+t^2k^4)^{-1}.
\en
Taking smoother initial data improves the fall off on the RHS
accordingly. In $x$ space, the convergence in the norm (3)
implies convergence both in $L^1$ and in $L^\infty$. For $L^1$,
we use Schwartz' inequality to get
\be
\| f \|_1 \leq C \| (1+|x|) f \|_2
\non
\en
and Plancherel to bound
\be
\| f \|_2 + \|xf\|_2 \leq C \|f\|
\non
\en
For $L^{\infty}$, we use $\|f\|_{\infty}\leq
\|{\hat f}\|_1 \leq C \|f\|$.

\vspace{3mm}

\no {\bf Proof.} We start by discussing the local
 existence of the solution. Thus, turn
(6) into an integral equation
\be
u(t)=e^{(t-1)\partial^2}f + \int_0^{t-1} ds
e^{s\partial^2}F(t-s)\equiv  u_f +N(u)
\en
where $F(\tau)=F(u(\tau),u'(\tau),u''(\tau))$ and $\partial =
\frac{d}{dx}$. We solve (10) using
the contraction mapping
principle by introducing the Banach space  of functions $u(x,t)$,
$t\in [1,L^2]$ with the norm  \be
\|u\|_L = \sup_{t\in [1,L^2]}\|u(t)\|.
\en
We shall show that $T(u) = u_f + N(u)$ maps the ball
$B_{f} = \{u | \| u - u_f \|_L \leq \|f\| \}$
into itself and is a contraction there, for
$\|f\|\leq\epsilon =
\epsilon (F,L)$ small.

We need to estimate $\|N(u)\|_L$. In this the use of momentum
space in the norm (3) is very convenient: we are indebted to
\cite{CEE} for this observation.
Thus, Taylor expand $F$ in (10) and take the Fourier transform:
\be
\hat{F}=\sum_{{\bf n}\in{\bf N}^3}a_{\bf n}\hat{u}^{*n_1}
\ast\widehat{u'}^{*n_2}\ast\widehat{u''}^{*n_3}
\en
where $\ast$ is the convolution. Inserting (12) into (10)
and using
the bound on the derivatives $|\widehat{u^{(l)}}(k)|\leq C\|u\|_L
(1+k^2)^{-1}$ , for $l=0,1,2$, we get
\be
\int_0^{t-1}ds e^{-s k^2}|(\hat{u}^{*n_1}
\ast\widehat{u'}^{*n_2}\ast\widehat{u''}^{*n_3})(k)|\leq
(C\|u\|_L)^{m}({1-e^{-(t-1) k^2}\over
k^2})
\non\\
\cdot\int(1+(k-p_1)^2)^{-1}  \dots (1+(p_m)^2)^{-1}dp_1\dots
dp_m\leq \hspace{20mm} \non \\
C_L[C\|u\|_L]^{m}(1+k^4)^{-1}.\hspace{30mm}
\en
with $m=n_1 + n_2 + n_3$.
The same estimate holds for the derivative with respect
 to $k$ and we get a
convergent bound on $\| N(u) \|_L$: since $ F$ is
analytic, $|a_{\bf n}|\leq (C_F)^m$, and hence
\be
\|N(u)\|_L\leq C_{L,F} \|u\|_L^2
\en
$(d_F > 0$ implies $m\geq2)$, provided $\|u\|_L$ is small enough.
This holds in $B_f$, if $\|f\|\leq\epsilon(F,L)$,  since
$\|u_f\|_L\leq C\|f\|$ and so $\|u\|_L\leq (C+1)\|f\|$. Thus
$T$ maps $ B_f$ into itself.

In the same way, we estimate
\be
\|N(u_1)-N(u_2)\|_L\leq C_F L^2(\|u_1\|_L+\|u_2\|_L) \|u_1
-u_2\|_L.
\en
Thus, for $\|f\|\leq\epsilon(F,L)$,
(10) has a unique solution in $B_{f}$
\be
u(\cdot,L^2)=e^{(L^2 - 1)\partial^2}f+v
\en
with
\be
\|v\|\leq C_{L,F}\|f\|^2
\en

Now we write
\be
f=A_0 f^*_0 +g_0
\en
with $A_0 =\hat{f}(0)$. Note that $\hat f^*_0(0)=1$, hence $\hat g_0
(0)=0$. The reason for this decomposition is that $R_0$ is a
contraction when it acts on functions $g$ with $\hat g (0) = 0$
(see (4)). Also,
\be  \|g_0\|=\|f-\hat{f}(0)f^*_0\|\leq C\|f\|.
\en
Then,
\be
Rf=Lu(L\cdot ,L^2)=R_0 f+Lv(L\cdot) =A_1f^*_0+g_1
\en
where $R_0$ is defined in (1),
\be
A_1=A_0+\hat{v}(0)
\en
and
\be
g_1=R_0g_0+Lv(L\cdot)-\hat{v}(0)f^*_0,
\en
so that $\hat g_1(0)=0$.
We have then the estimates, using (17),
\be
|A_1-A_0|\leq C_{L,F}\|f\|^2
\\
\|Lv(L\cdot)-\hat{v}(0)f^*_0\|\leq C_{L,F}\|f\|^2
\en
and, combining with (4), we get
\be
\| g_1 \|\leq
CL^{-1}\| g_0 \|+C_{L,F}\|f\|^2\leq L^{-(1-\delta)}\|f\|
\en
for $\|f\|\leq\epsilon (L,F)$ and $L$ large (depending on $\delta$).

The proof is now completed by iterating this procedure. Set
\be
f_n\equiv R_{L^n ,F}f=A_n f^*_0 +g_n
\en
So, with $ v$ above equal $v_0$,
\be
A_{n+1} = A_n + \hat{v}_n (0) \\
g_{n+1} = R_0 g_n + L v_n (L\cdot) - \hat{v}_n (0)f^*_0
\en
and assume inductively that
\be
| A_n | \leq (C-L^{-n}) \|f\| \hspace{5mm},\hspace{5mm} \| g_n
\|\leq  CL^{-(1-\delta)n}\|f\|,
\en
so that $\|f_n\| \leq C \|f\|$.
Then, repeating the above analysis and noting that in (14) and
(15), $C_F$ is replaced by $L^{-nd_F}C_F$ (see (2.5)), we get,
instead of (17), (23), (25),
\be
\|v_n \| \leq C_{L,F}L^{-nd_F} \|f\|^2
\en
\be
|A_{n+1}-A_n| \leq C_{L,F}L^{-nd_F} \|f\|^2
\\
\|g_{n+1}\| \leq C L^{-1}\| g_{n}\| + C_{L,F} L^{-nd_F}\|f\|^2.
\en
Thus, (29) iterates ($d_F \geq 1$) and we get that $A_n\rightarrow A$ with
$|A_n - A| \leq C_{L,F} L^{-nd_F} \|f|\|^2 \leq CL^{-nd_F} \|f\|$
for $\|f\| $ small,
\be
|A-\hat{f}(0)|=|A-A_0| \leq C_{L,F}\|f\|^2
\en
and (see (2.7), (26)), for $t=L^{2n}$,
\be
\| u(\cdot t^{1/2},t)-At^{-{1\over 2}}
f^*_0(\cdot)\| \leq Ct^{-1+\ha\delta} \|f\|.
\en
It is trivial to extend this bound to $t=\tau L^{2n}$,
 with $1\leq \tau \leq L^2$ by replacing everywhere $L^2$
by $\tau L^2$. This proves the claim. \hfill $\Box$

\vspace{3mm}

\no {\bf B. The marginal case}

\vspace{3mm}

Let us now consider the marginal cases, i.e.
\be
F=-u^3+G(u,u',u'')
\en
or
\be
F=2uu'+H(u,u',u'') = (u^2)' +H(u,u',u'')
\en
where $G$ and $H$ are irrelevant. (36) is
just the Burgers equation
with a perturbation. Since the discussion of
that equation uses
some ideas introduced in the next section,
we shall discuss it at
the end of Section 4. In (35), we need the
negative sign: for $G=0$
and a positive sign, any non zero initial
data leads to a solution
that blows up in a finite time \cite{L,CEE}.
However, we do   not have
to assume that  the initial value is pointwise
positive; rather we
 want $f$ to be near a
small multiple of  $f^*$. Although a general
 analytic irrelevant $G$
could be treated,  we assume for simplicity
that the Taylor
expansion of $G$ starts  at degree 4 or higher.
 Without this
assumption, in the proof below,   we would
need to consider first a
"crossover" time during which the  $G$ term
 would dominate $u^3$.
We will therefore scale $u$ by   $\lambda^\ha$, where $\lambda$
is chosen so that $\hat{u}(\cdot,1)(0) =1$, and consider the
equation
\be
\dot{u}=u''-\lambda u^3 +G_\lambda (u,u',u'')
\en
with $G_\lambda(z)=\lambda^{-\ha}G(\lambda^\ha z)$ (which is of
order $\lambda^{3/2}$ for $\lambda$ small) and initial data
\be
f=f^*_0+g\; ,\;\;\hat{g}(0)=0.
\en
We have then

\vspace{3mm}

\no {\bf Theorem 2.} {\it For any $\delta>0$ there exist $\lambda_0\;
,\;\epsilon>0$ such that for $0<\lambda\leq\lambda_0$ and
$\|g\| \leq  \epsilon$, we have
\be
\lim_{t\rightarrow\infty} t^{1\over 2}(\log t)^{1 -\delta}
\|u(\cdot t^{\ha},t)-
(\frac{\lambda}{2\sqrt3\pi}t\log t)^{-\ha}
f^*_0(\cdot)\|=0
\en}

\vspace{3mm}

\no {\bf Proof.} We proceed as in the proof of Theorem 1. We use
(10) and we get, for $\lambda $ small, instead of (14),
\be
\|N(u) \|_L \leq C_{L,G} \lambda
\en
($\|u\|_L$ is of order one here). This, with the analogue of
(15), proves local existence, leading
to  \be
Rf=R_0f+Lv(L\cdot)
\non
\en
with
\be
\|v\| \leq C_{L,G}\lambda
\en

Since $u^3$ is marginal, we have to handle rather explicitely
its main effect on $u$ which will lead to the logarithmic
correction in (39). Let, in (10), $N(u)=-\lambda N_3(u)+N_G(u)$
corresponding to
the two terms in  (37). So,
\be
N_3(u) (t) =\int_0^{t-1}ds
e^{s\partial^2}(u(t-s))^3
\non
\en
We have
\be
\|N_G(u)\|_L \leq C_{L,G}\lambda^{3/2}
\en
Denote by $u^*_A$ the solution of (37) with $G_{\lambda} = 0$ and
$f=Af^*_0$, and by
$u_A$  the solution of (37) with $f= Af^*_0 + g$.
So,
\be
 u^*_A (t) =Ae^{(t-1)\partial^2}f^*_0 - \lambda N_3(u^*_A) (t)
\en
\be
 u_A (t) =Ae^{(t-1)\partial^2}f^*_0 + e^{(t-1)\partial^2}g
- \lambda N_3(u_A) (t) + N_G (u_A) (t)
\en
Here $A=1$, but we shall use below the following bound,
for $|A|\leq1$ and $\|g\| \leq \epsilon$:
\be
\|N_3(u_A)-N_3(u^*_A)\|_L \leq C_L ( A^2\|g\|
+ \|g\|^3)+ C_{L,G} \lambda^{3\over 2}.
\en
To prove (45), first show, like in (15),
 but with a cubic nonlinearity, that
\be
\|N_3(u_A)-N_3(u^*_A)\|_L \leq C_L
(\|u_A\|_L + \|u_A^*\|_L)^2 \|u_A -u_A^*\|_L
+C_{L,G} \lambda^{3\over 2}
\non
\en
Then, use (42), (43), (44) to show
\be
\|u_A -u_A^*\|_L \leq C_L \|g\| + C_{L,G}\lambda^{3 \over 2}
+\lambda \|N_3(u_A)-N_3(u^*_A)\|_L
\non
\en
Finally, use $\|u_A^*\|_L \leq C_L |A|$ and
$\|u_A\|_L \leq C_L( |A| + \|g\|)$,
which follows from (43), (44); (45) follows for $\lambda$ small.

If we define
\be
v^* =\int_0^{L^2 -1}ds e^{s\partial^2}(e^{(L^2-s-1)\partial^2}f^*_0)^3
\non
\en
we get, inserting (43) in $N_3(u^*_A)$, and using the bound
$\|N_3(u^*_A)\|_L \leq C_L  |A|^3$,
\be
\|N_3(u^*_A) (L^2) - A^3 v^*\| \leq C_L \lambda A^5
\en
for $|A| \leq 1$. Now, write
\be
v=-\lambda v^* +w
\en
where, using (42), (45),(46), for $A=1$ and $\|g\|\leq \epsilon$,
\be
\|w\|\leq C_{L,G}\lambda(\|g\|+\lambda^\ha).
\en
((46) gives a term of order $\lambda^2$ here). Again, we write
\be
f_1\equiv Rf =A_1f^*_0 +g_1
\en
with $A_1=1+\hat{v}(0)$ ($A_0=1$ here, by (38)) and
$g_1$ given by (22). Denote $\hat{v}^*(0)=\beta$. We get from
(48)
\be
|A_1-1+\lambda \beta|=|\hat w(0)| & \leq  &
C_{L,G}\lambda (\|g\|+\lambda^\ha)
\en
and, from (4), (41),
\be
\|g_1\|& \leq &
CL^{-1}\|g\|+C_{L,G}\lambda
\en

The iteration is as follows: the $G$ term will run down
with a factor $L^{-nd_G}$, which improves the bound (42)
and the last term in (45), but we shall see that, unlike the
situation of Theorem 1, $A_n$ will go to zero and we have to keep track
of the correction of order $A^5$ in (46). We use
(26), (27), (28) and we have
\be
v_n=-\lambda A^3_nv^*+w_n
\en
with, using again (42), (45), (46),
\be
\|w_n\| \leq C_{L,G}\lambda (A^2_n \|g_n\|
+ \|g_n\|^3+\lambda^{\ha}L^{-nd_G} +\lambda
A_n^5).
\en
So,
\be
|A_{n+1}-A_n+\lambda \beta A^3_n| & \leq  & C_{L,G}\lambda
(A_n^2 \|g_n\| + \|g_n\|^3 +  \lambda^\ha L^{-nd_G} +\lambda A_n^5)
\en
Using the fact that $\beta>0$ (see (58) below),
we see that $A_n$ decreases
and so, using (28), (4), (52), (53), $\|v^*\|\leq C$,
and $C_{L,G} \lambda A_n^2 \leq L^{-1}$,
for $\lambda$ small, we get
\be
\|g_{n+1}\| & \leq & CL^{-1}\|g_n\|+C_{L,G}\lambda( A_n^3
+ \|g_n\|^3 +
\lambda^{1\over 2} L^{-nd_G}
 + \lambda A_n^5).
\en

The iteration of $A_n , g_n$, leads to
\be
 A^2_n=(2\lambda \beta n +b_n)^{-1}
\en
\be
\|g_n\| &\leq &C_{L,G} n^{-{3\over 2}}
\en
for $n\geq 1$, with $|b_{n+1} -b_n| \leq C_{L,G} n^{-\ha}$,
which gives
$|b_n| \leq C_{L,G}\sqrt{n} $. Indeed, the leading term
in (55) is $A_n^3$ and, in (54), $A^2_n \|g_n\| + A^5_n$. We get
$A_n = (2\lambda \beta n)^{-\ha} + {\cal O} (n^{-1})$.
Finally, we compute
\be
\beta={1\over 4\pi^2}\int_0^{L^2 -1}ds\int_{{\bf R}^2}dpdq
e^{(s-L^2)(p^2+q^2+(p-q)^2)}={\log L\over 2\sqrt3\pi}.
\en
Using $2n \log L =\log t$, the claim follows as in Theorem 1.\hfill $\Box$

\vs{1cm}

\addtocounter{equation}{-58}
\no  {\Large\bf 4. Non-Gaussian fixed point: stability of a front }

\vs{1 cm}

If equation (2.1) has other (constant) stationary solutions
than $u=0$, we may study the problem where $\lim_{x\rightarrow
\pm\infty}u(x,1)$ takes two different values. This corresponds to
a $front$ in the initial data. One may then inquire about the
stability of this front and the universal features of the long
time solution in such a situation. These problems are discussed
in general in \cite{B,CE1} and, for coupled equations
related to the equations discussed here, see \cite{CEE,CE2,BK}.
Here, the restriction to one space dimension is
essential.

An example of such an equation is the nonlinear diffusion equation
\be
\dot{u}=\p((1+a(u,\p u))\p u)\;\; ,\;\; u(x,1)= \phi (x) \;\; ,\;\;
\lim_{x\rightarrow   \pm\infty} \phi (x)=u_\pm
\en
where we denote $\p = {d\over dx}$. This is a special case of (2.1),
with $F=au'' + {\p a \over \p u} (u')^2 + {\p a \over \p u'}u'u''$. If $a$
is analytic and if $a(0,0)=0$, $F$ is irrelevant.
(1) has $u={\rm const.}$
as a stationary solution.
\vspace{3mm}

\no {\bf Remark 1}. The most general such equation in one variable
we could deal with is
\be
\dot{u}=(1+a(u,u',u''))u''+ b(u,u',u'')(u')^2
\en
with $a$ and $b$ analytic as before. See Remark 2
below.

\vspace{3mm}

To understand the problem in the RG setup, let us first consider the
trivial case $a=0$. This is of course exactly soluble. We have
\be
u(\sqrt{t}x,t+1)={1\over \sqrt{4\pi}}\int e^{-{1\over 4}(y-x)^2}
\phi(\sqrt{t}y)dy\non\\
\rightarrow_{t\rightarrow\infty}
u_{-}+(u_{+}-u_{-})e(x)\equiv
\phi^*_0(x)
\en
where $e(x)=\int_{-\infty}^x e^{-{1\over 4}y^2}{dy\over\sqrt{4\pi}}$.
In RG terminology, we have the "Gaussian" fixed point $\phi^*_0 $
corresponding to the $u_\pm$ boundary condition problem. It is
easy to check that $\phi^*_0 $ is a fixed point for the map
\be
R_L\phi=u(L\cdot,L^2)\;\; ,\;\;u(\cdot,1)=\phi
\en
Note the absence of the multiplicative $L$ factor
in $R_L$ (in (2.3), we
have $\alpha = 0$): the  initial data are not
normalizable by their
integral and $u(x,t)$ does not, in general,
go to zero as $t$
goes to infinity.

The stability of this fixed point is also easy to understand. We
write $\phi=\phi^*_0+f$, where $f(\pm\infty)=0$. Then $\dot{v}
= v''$, with $v (x,1) = f(x)$, and the analysis
 of the previous section
applies.  Thus we expect
\be
u(x,t)\sim\phi^*_0({x\over\sqrt{t}})+
\frac{\hat{f}(0)}{\sqrt{t}} f^*_0({x\over\sqrt{t}})
+{\cal O}(t^{\delta-1})
\en
This is the asymptotics we wish to prove for (1), except that
$\phi^*_0$ and $f^*_0$ are replaced by non trivial fixed points.
We will prove a theorem for small data, so $u_\pm$ will be taken
small. What we do below is an (elementary) example of
what is called, in the theory of critical phenomena, the "epsilon
expansion". Before stating the precise results, let
us do the  heuristics.

Upon scaling $u_L(x,t)=u(Lx,L^2t) $, $u_L$ satisfies (1)
with $a$ replaced by
$a_L(u,u')=a(u,L^{-1}u')$. Thus, we will search for a
fixed point for the RG (4) with $u$ satisfying the equation
obtained in the limit $L \rightarrow \infty$:
\be
\dot{u}=\p((1+a^*(u))\p u)\;\;,\;\;u(x,1)=\phi^*(x)
\en
with $\phi^*(\pm\infty)=u_\pm$ and $a^*(u)=a(u,0)$. We will
find the fixed point $\phi^*$ by finding a scale invariant
solution to (6):
\be
u(x,t)=\phi^*({x\over\sqrt{t}}).
\en
We get (replacing ${x\over\sqrt{t}}$ by $x$)
\be
\p((1+a^*(\phi^*))\p\phi^*) +\ha x\p\phi^*=0
\en
and we look for a solution
\be
\phi^*=\phi^*_0+\psi
\en
with $\psi(\pm\infty)=0$ and $\phi^*_0$ is the Gaussian solution
(3). Thus, we get for $\psi$ the equation
\be
A\psi=\p(a^*(\phi^*_0+\psi)\p(\phi^*_0+\psi))
\en
with
\be
A=-\p^2-\ha x\p.
\en
Let us also denote
\be
\psi_0=\p(a^*(\phi^*_0)\p\phi^*_0).
\en
We shall discuss the properties of $A^{-1}$ later.
We solve the fixed point equation (10) in the space
 of $C^N$ functions
equipped with the norm
\be
\|\psi\|_N=\max_{0 \leq m\leq N}\sup_{x}|\p^m\psi(x)|e^{{x^2\over 8}}.
\en
Let the degree of $a^*$ (i.e. $a^*(u)={\cal O}(u^d)$ for
 $u$ small) be $d>0$.
Then  we have the

\vspace{3mm}

\no{\bf Proposition}. {\it Let $a:{\bf C}^2\rightarrow{\bf C}$ in
$(1)$ be analytic in a neighbourhood of $0$ and let
$N$ be an integer.
Then there exists an $\epsilon>0$ such that, for
$|u_{\pm} | \leq \epsilon$
in $(3)$, $(10)$ has a unique   solution with
\be
\|\psi-\psi_0\|_N\leq C\epsilon^{2d+1}.
\non
\en}

\vspace{3mm}

\no The proof of the proposition will be given after the one of
 Theorem 3
below. We will also see that $\psi_0$ can be written down
 explicitely and is
${\cal O}(  \epsilon^{d+1})$ in the norm (13).

\vspace{2mm}

Given $\phi^*$, we return to equation (1), and write
\be
u(x,t)=\phi^*({x\over\sqrt{t}})+v(x,t)\equiv u^*(x,t) + v (x,t) .
\en
Because of (6, 7), $v$ satisfies the equation
\be
\dot{v}=\p((1+a)\p v+(a-a^*(u^*))\p u^*)\;\; ,\;\; v(x,1)=f(x)
\en
with $\lim_{x\rightarrow\pm\infty}f(x)=0$. Now the RG is as in
Section 3:
\be
R_L f=Lv(L\cdot,L^2)\equiv v_L(\cdot ,1)
\en
However, the fixed point will not be the Gaussian $f^*_0$
of that Section, but a different one. We find it by
taking the scaling limit: $v_L$ satisfies the equation
\be
\dot{v_L}=\p((1+a_L)\p v_L
+L(a_L-a^*(u^*))\p u^*)
\en
with $a_L=a(u^*+L^{-1}v_L,L^{-1}\p(u^*+L^{-1}v_L))$,
and therefore we look for a fixed point satisfying the linear
equation obtained from (17) by letting $L\rightarrow\infty$:
\be
\dot{v}^*=\p(\p((1+a^*(u^*))v^*) + {{\p a}\over\p u'}(u^*,0) (\p u^*)^2).
\en
This is solved by setting
\be
v^*(x,t)={1\over\sqrt{t}}f^*({x\over\sqrt{t}})
\en
whence, $\p(\p((1+a^*(\phi^*))f^*)+\ha x f^* + h)=0$
where $h={\p a\over\p u'}(\phi^*,0) (\p\phi^*)^2$. We solve this equation
by integrating once (we shall look for an
integrable $f^*$, so the constant
of integration is zero) and then solving a first order equation:
\be
f^*(x)=(\int_0^{x} h(y) F(y) dy +{\cal N})
e^{-\ha\int_0^{x}{y+2\p a^*(\phi^*(y))\over
1+a^*(\phi^*(y))}dy}.  \en
where $ F(x)=-(1+a^*(\phi^* (x))^{-1}
e^{+\ha\int_0^{x}{y+2\p a^*(\phi^*(y))\over
1+a^*(\phi^*(y))}dy}$.
For future convenience, we normalize this by choosing $\cal N$
such that $\int f^* = {\hat f^*}(0)=1$.  Note that $f^*$ is, for
 $\phi^*$ small, a small
perturbation of $f^*_0$. In particular, $f^*$ is smooth and decays rapidly
at infinity.

\vspace{2mm}

\no{\bf Remark 2.} This is the only place where (1) is
simpler than
(2). For (2), (18) is replaced by
\be
\dot{v}^*=  (1+a^*(u^*))\p^2 v^*+2b^*(u^*)\p u^*\p v^*
+({\p a^* \over \p u^*}(u^*)(\p^2 u^*) +\non\\
{\p b^* \over \p u^*}(u^*)(\p u^*)^2)v^*
 +({\p a \over \p u'}(u^*,0,0)(\p^2 u^*) + {\p b
\over \p u'} (u^*,0,0)(\p u^*)^2)(\p u^*)
\en
with $a^*(u)=a(u,0,0)$ and $b^*$ similarily. $v^*$ will
 not be
as explicit as (19,20) and $1\over\sqrt{t}$ in front of $f^*$ in (19)
will be replaced by $t^{(-\ha+{\cal O}(\epsilon))}$ since,
while $\int v^*$ is constant for the solution of (18) (integrate
both sides), it is not conserved by (21); see \cite{G1,G2}
for a discussion of a similar effect.

\vspace{2mm}

Before stating the main result of this section, we need to
specify the space of initial data $\phi$ of (1), i.e. the $f$
of (15). We can not directly use the norms (3.3) introduced in
Section 3. The reason is that the function $\phi^*$ in (9)
involves the Gaussian fixed point (3) which has no falloff
at infinity. In particular it is not in the Banach space we
used before. However, the only way $u^*$ enters in (15) is
as $\p u^*$ that falls off like a gaussian, or
by multiplying functions that we expect to fall off at infinity.
As a consequence of the latter possibility, we may not use
pure momentum space bounds as in (3.13):
the convolutions would
then involve $\widehat{u^*}$ which is too
 singular ($|\hat{\phi}^*_0| \simeq
|k|^{-1}$ for small $k$). Instead, we introduce
 a norm that encodes more
sharply than (3.3) both the long and short
distance properties of the
solution.

Thus, let $\chi$ be a non negative $C^\infty$ function on
$\bf R$ with compact support on the interval $(-1,1)$,
such that its translates by $\bf Z$, $\chi_n=\chi(\cdot-n)$,
form a partition of unity on $\bf R$. For $f\in C^2$, we
then introduce the norm
\be
\|f\|=\sup_{n\in{\bf Z},k\in{\bf R},i\leq 2}(1+n^4)(1+k^2)
|\widehat{\chi_n \p^if}(k)|.
\en
Roughly, $\|f\|<\infty$ means that $f$ falls off at least as
$x^{-4}$ at infinity and $\hat{f}(k)$ as $k^{-2}$. Note that the
derivatives of $\phi^*$ and $f^*$ and its
 derivatives have a finite norm.
Comparing with (3.3), we have $k^2$ instead
of $k^4$ but two derivatives
act on $f$. We used $k^2$ to do the
 convolutions in (3.13). The  $n^4$ could
be changed to anything increasing not faster
than $e^{{n^2\over 8}}$
(coming from (13)).   These norms for different
 choices of $\chi$ may be
shown to  be equivalent.

\vspace{3mm}

\no{\bf Theorem 3.} {\it Let $a:{\bf C}^2\rightarrow {\bf C}$
be analytic in a neighbourhood of $0$ with positive degree
and fix a $\delta>0$. There is an $\epsilon>0$
such that for $|u_\pm|$, $\|f\|\leq\epsilon$, the
equation
\be
\dot{u}=\p(1+a(u,\p u)\p u)\;\;,\;\; u(x,1)=\phi^*(x)+f(x)
\en
has a unique solution, satisfying
\be
\lim_{t\rightarrow\infty}t^{1-\delta}\|u(\sqrt{t}\cdot,t)-
\phi^*(\cdot)-{\hat{f}(0)
\over\sqrt{t}}f^*(\cdot)\|=0
\en
where $f^*$ is given in $\rm(20)$ and $\phi^*$ in
the Proposition.   }

\vspace{3mm}

\no{\bf Remark 3.} The convergence in the norm (22) again implies
convergence in $L^1$ and in $L^{\infty}$,
see equations (39) and (40)
below, applied to $i=0$. Equation (24) means that $u(x,t)$ behaves,
for $|x|\leq C$ and $t\rightarrow\infty$ as $\phi^*(0) + A t^{-\ha} +
{\cal O}(t^{\delta-1})$, where $A=(\phi^*)'(0)+\hat{f}(0)
f^*(0)$. Thus, locally, the solution goes to a constant
i.e. a stationary solution of (1), with a diffusive
correction. This solution is "selected" by the boundary condition
at infinity, $u_{\pm}$, and by $a$.

\no{\bf Remark 4.} Since $\psi$ in (9)
satisfies $\| \psi \| \leq  C \epsilon^2$
(this
follows easily from the Proposition) we can replace
 $\phi^*$ in (23) by
$\phi^*_0$, given by (3), which makes
the hypothesis more explicit.

\vspace{3mm}

\no{\bf Proof}. We consider the equation (15) and the RG (16).
It is convenient to separate the $f^*$ piece from $f$. We write
\be
f=\hat{f}(0)f^*+g
\en
and correspondingly
\be
v=v^*+w
\en
with
\be
v^*(x,t)={\hat{f}(0)\over\sqrt{t}}f^*({x\over\sqrt{t}}).
\en
First, we have $|\hat{f}(0)|\leq \sum_{n\in{\bf Z}}
|\widehat{\chi_n f}(0)| \leq
\sum_{n\in{\bf Z}} \frac{\|f\|}{1+n^4}\leq C\|f\|  $
and $\|f^*\|\leq C$,
which implies
\be
\|g\|\leq C\|f\| \leq C \epsilon.
\en
The equation we finally study
is the one
satisfied by $w$:
\be
\dot{w}=\p^2 w+K\;\; ,\;\; w(x,1)=g(x)
\en
where ${\hat g}(0)=0$ (see (25) and the normalisation in (20))
and where (see (15), (18))
\be
K=\p[(a -a^*(u^*))\p v^*+ a\p w  +(a-a^*(u^*) -
{{\p a}\over\p u'}(u^*,0) (\p u^*)))\p u^*-v^*\p a^*(u^*)]
\en
with $a=a(u,u')$, $u=u^*+v^*+w$.
The explicit form of $K$
is not important for us, all we need to know is that it is
given as a convergent power series in $u^*$, $v^*$,
 $w$ and their
derivatives up to second order, has no linear term
and no
term which is only a power of $u^*$. We shall also use some scaling
properties of $K$ discussed below.

In RG language, we need to control (see(16))
\be
L^nw(L^nx,L^{2n})=R_{L^n,K}g=R_{L,K_{n-1}}\dots R_{L,K}g
\en
where now $K_{n+1}$ is obtained from $K_n$ by replacing (see (16), (4))
\be
\p^iv^*\rightarrow L^{-1-i}\p^iv^*\; ,\;
\p^iw\rightarrow L^{-1-i}\p^iw\; ,
\p^iu^*\rightarrow L^{-i}\p^iu^* .
\en
To see the behaviour of the terms in (30)
 under this scaling,
define $d=n_{\p} + n_{v^*} + n_w -3$
where $n_{v^*}, n_w$ count the number of $v^*,w$ factors and their
derivatives,
while $n_{\p}$ is the total number of derivatives.
We say that a term in (30) is irrelevant if $d>0$ and marginal if $d=0$.
A term with $d<0$ would be relevant, but these terms have been included
in the equations for $u^*$ and $v^*$. To check this, notice that
 each term in (30) has either more than
two derivatives or two derivatives and
at least one $v^*$ or $w$ factor, so that $d\geq 0$.
In fact, the only marginal terms
in (30) are of the form  $(u^*)^l\p^2 w$ or
$(u^*)^l\p u^*\p w$. To check that there are
no terms like that with $w$ replaced by $v$, observe that such terms are
cancelled because of the $a-a^*$ in front of $\p v^*$ and because
of the subtraction between the two last terms in (30). A term like
$\p{{\p a}\over\p u'}(u^*,0) (\p u^*)^2$ is also marginal, but it is
cancelled by the subtraction in the factor multiplying $\p u^*$ in (30).

 Eventually, using ${\hat g}(0)=0$, we
will show that $\| R_{L^n,K}g \|$ goes to zero like $L^{-(1-\delta)n}$.
Using
(14), (26), (27), (31), this will prove (24).
We split the proof again into three steps: the
proof of local existence,
the control of $R$ and the iteration of $R$.

\vspace{2mm}

\no (a) {\it Local existence}. We have to solve the
 integral equation
\be
w(t)=e^{(t-1)\p^2}g+\int_0^{t-1}ds e^{s\p^2}K(t-s)\non
\en
(with obvious notation) using the contraction mapping
 principle with
the norm
\be
\|w\|_L=\sup_{t\in [1,L^2]}\|w(\cdot ,t)\|.
\en
As in Theorem 1, we expand $K$ as a power series and estimate
a generic term
\be
\alpha(x)=\int_0^{t-1}ds\int dy e^{s\p^2}(x-y)u^*(y)^l F(y)\non
\en
where $F(y)$ is a (non empty) product of the functions
  $v^*$, $w$
and their derivatives, and derivatives of $u^*$, and $l\geq 0$.
We supressed the $t-s$ variable in $u^*$ and $F$.
We localize the $y$ variable:
\be
\alpha(x)=\sum_{m\in{\bf Z}}\int_0^{t-1}ds\int dy
e^{s\p^2}(x-y)\chi_m(y)u^*(y)^l F(y)\equiv\sum_{m\in{\bf Z}}
\alpha_m(x).
\en
We want to bound
\be
\beta_{mn}=\sup_{k} |(1+k^2)\widehat{\chi_n\p^i\alpha_m}(k)|.
\en
We distinguish between $|n-m|\geq 2$ and $|n-m|<2$.

\vspace{2mm}

\no (A) Let first $|m-n|\geq 2$. Then $\chi_m$ and $\chi_n$ have
disjoint supports and $e^{s\p^2}(x-y)$ is smooth uniformly in $s$.
We write
\be
\beta_{mn}=\sup_{k}|\int dx e^{-ikx}\int_0^{t-1}ds\int dy
(1-\p_x^2)(\chi_n(x)\p_x^i e^{s\p^2}(x-y))\chi_m(y)u^*(y)^l F(y)|
\en
and estimate the various factors on the RHS.

First, we use, for $j\leq 4$,
\be
|\p^{j}e^{s\p^2}(x-y)|\leq Ce^{-|m-n|}
\en
on the support of $\chi_m$ and $\chi_n$. To bound $F(y)$, we
need the bounds (for $0\leq i\leq 2$)
\be
\|\p^i u^{*}\|_\infty\; ,\;
\|\p^i v^{*}\|_\infty \leq C\epsilon
\en
which follow from (3), the Proposition and (20,27,28), and the bound
\be
|\p^i w(x)|=\sum_n |\chi_n(x)\p^i w(x)|
\leq\sum_n\int dk|\widehat{\chi_n\p^i w}(k)|\non\\ \leq
\sum_n (1+n^4)^{-1}\int dk (1+k^2)^{-1}\|w\|\leq C\|w\|
\en
To extract the factor $(1+m^4)^{-1}$, we show that
\be
\int|\chi_m(y)\p^iw(y)|dy\leq{C\|w\|\over 1+m^4}
\en
and similarily for $v^*$ and $u^*$ (with $i>0$
for the latter)
with $\|w\|$ replaced by $\epsilon$.

Indeed, let $\phi_m\in C^\infty_0({\bf R})$ be
such that
$\phi_m\chi_m=\chi_m$. Then
\be
\int|\chi_m\p^iw|= \int|\chi_m\phi_m\p^iw|\leq
\left(\int\phi^2_m\right)^\ha
\non\\
\cdot \left(\int|\chi_m\p^iw|^2 \right)^\ha
\leq C \left(\int|\widehat{\chi_m\p^iw}|^2 \right)^\ha
\leq {C\|w\|\over
1+m^4}   \en
For $v^*$ and $u^*$ (with $i>0$ for the latter) the bound holds
with $\|w\|$ replaced by $\epsilon$ due to the explicit
expressions and bounds (3), (20) and the Proposition.

Now we bound (36): the $x$ integral is controlled
by $\chi_n (x)$ or its
derivatives, the $s$ integral is less than $L^2$
and we use (38) and (39)
for $u^*(y)^l$ and all factors in $F(y)$ except
 one, for which we use
(40):
\be
\beta_{nm}\leq L^2 C^{l+M+N}
e^{-|m-n|}(1+m^4)^{-1}\epsilon^{l+M}
\|w\|_L^N
\en
with $M$ the total number of factors
of $v^*$ and its
derivatives and of derivatives of $u^*$ in $F$,
and $N$ similarily for
$w$.  We have $l+M+N \geq 2$.

\vspace{2mm}

\no (B) Let now $|m-n|<2$. The difficulty is that
we do not have (37) for
$s$ close to zero. But we do not have to control
 a sum over $m \in {\bf
Z}$, and we can use Fourier transforms. Let us
denote $\phi_mu^{*l}$ by $f_m$
where $\phi_m$ is as in (A). Then  \be
\widehat{\chi_n\p^i\alpha_m}(k)=\int
dsdpdq\hat{\chi}_n(k-p)(ip)^i
e^{-sp^2}\hat{f}_m(p-q)\widehat{\chi_mF}(q).
\en
Let us consider the various factors on the RHS.
Since $\chi$ is
$C^\infty$ with compact support, we have
\be
|\hat{\chi}_n(k-p)|= |e^{-i(k-p)n}\hat{\chi}(k-p)|\leq C_l
(1+|k-p|^l)^{-1}
\en
for any $l$.

For $\hat{f}_m$, note that
\be
\int|(1+(-\p^2)^r)\phi_m u^{*l}(x)|dx\leq C_r(C\epsilon)^l
\en
for all $r$, whence
\be
|\hat{f}_m(p-q)|\leq C_r(C\epsilon)^l(1+(p-q)^{2r})^{-1}.
\en
Also, $\int ds |p|^je^{-sp^2}\leq CL^2$ if $j\leq 2$, so,
provided we can show
\be
|\widehat{\chi_mF}(q)|\leq C^{N+M}\epsilon^M\|w\|^N(1+m^4)^{-1}
(1+q^2)^{-1}
\en
we can perform the convolutions in (43) to get
\be
\beta_{mn}\leq L^2 C^{l+N+M}(1+m^4)^{-1}\epsilon^{l+M}\|w\|^N_L.
\en
Using (42),(48), and the contraction mapping
principle as in the proof of Theorem 1, we find a solution $w (x,t)$
such that
\be  \|w(\cdot,L^2)-e^{(L^2-1)\p^2}g(\cdot)\|
\leq C_{L,a}\epsilon(\epsilon+\|g\|).
\en
The first term comes from $N=0, l+M\geq 2$ and
the second from $l+M\geq 1, N\geq 1$. To apply the
contraction mapping principle,
we used
$\| e^{(t-1) \p^2} g \|_L \leq C \|
g \|$ which is proven like (52) below.

To prove (47), use
$$|\widehat{\chi_m\p^iw}|\leq (1+m^4)^{-1}(1+k^2)^{-1}\|w\|
$$
and
$$|\widehat{\p^iw}(k)|=|\sum_n\widehat{\chi_n\p^iw}(k)|\leq
{C\|w\|\over1+k^2}$$
together with similar bounds for $v^*$ and the
derivatives of $u^*$ and
perform the  convolutions as in equation (3.13).

\vspace{2mm}

\no (b) {\it Contraction}. Let us now turn to the RG (29), i.e.
\be
Rg=Lw(L\cdot,L^2).
\en
Note that because of the derivative in $K$
(see (30)), $R$ preserves
the condition $\hat{g}(0)=0$ i.e., by (29),
$\frac{d}{dt} \int w (x,t) dx = 0$ and so,
$\int (Rg) (x) dx = \int g (x) dx = 0$. We will show that
\be
\|Rg\|\leq {C\over L}\|g\|+C_{L,a}\epsilon(\epsilon+\|g\|).
\en
Indeed, by (49), all we need is to show
\be
\|R_0g\|=\|L(e^{(L^2-1)\p^2}g)(L\cdot)\|\leq{C\over L}\|g\|.
\en
It is now easier to work in the $x$ representation. We have
\be
R_0g=\int G(x,y)g(y)dy
\en
with
\be
G(x,y)=(4\pi(1-L^{-2}))^{-1}e^{-{1\over 4}(1-L^{-2})^{-1}
(x-L^{-1}y)^2}.
\en
We need to bound
\be
\|R_0g\|=\sup_{k,n,i \leq 2}(1+n^4)|\int dx e^{ikx}\int dy
(1-\p^2_x)(\chi_n(x)\p^i_xG(x,y))g(y)|.
\en

We consider first $|n|\geq C\log L$. Then
\be
|\dots|\leq\sum_m\int dxdy|(1-\p^2_x)(\chi_n(x)\p^i_xG(x,y))
\chi_m(y)g(y)|
\non\\
\leq C\sum_m e^{-|n-L^{-1}m|}(1+m^4)^{-1}\|g\|
\leq CL^{-3}{\|g\|\over (1+n^4)}
\en
for $L$ large enough. We used (41)
(with $\p^i w$ replaced by $g$) and (54). To control the sum,
we used $|n|\geq C\log L$ for $|n-L^{-1}m|\geq {|n|\over 2}$, and
 $(1+m^4)^{-1}$ for
$|n-L^{-1}m|\leq {|n|\over 2}$, which implies $m\geq L{|n|\over 2}$.

For $|n|\leq C\log L$, we subtract $G(x,0)$
from $G(x,y)$ in (55).
We may do this for free, since $\int g(y)dy=0$. Then,
\be
|\dots|\leq\sum_m \int dxdy|(1-\p^2_x)(\chi_n(x)\p^i_x(G(x,y)-
G(x,0)))
\chi_m(y)g(y)| .
\en
For the terms with $|m|\leq L$, we use
\be
|\p^j(G(x,y)-G(x,0))|\leq CL^{-1}e^{-|n|}(1+|m|)
\en
and thus, using (41) again,
\be
\sum_{|m|\leq L}\leq CL^{-1}\sum_{|m|\leq L}
e^{-|n|}{\|g\|\over 1+|m|^3}
\leq CL^{-1}{\|g\|\over 1+n^4} .
\en
For the terms with $|m|\geq L$,
the subtraction in (57) is not necessary, (41) suffices.
\be
\sum_{|m|\geq L}\leq C\sum_{|m|\geq L}
{\|g\|\over 1+m^4}
\leq C L^{-3} \|g\|
\leq CL^{-1}{\|g\|\over 1+n^4}.
\en
for $L$ large, since $|n|\leq C \log L$. Hence, (52) follows.

\vspace{2mm}

\no (c) {\it Iteration}. To conclude the proof,
 we need to iterate
$R$. Going back to (31), we study $R_{L,K_n}$.
 From the expression (30)
for $K$ and from (32), we see that the only change
to the $n=1$ analysis
above will be a change in (51):
\be
\|R_{L,K_n}g_n\|\leq {C\over L}\|g_n\|+C_{L,a}
\epsilon( L^{-n} \epsilon
+\|g_n\|).
\en
Indeed, all the terms in (30) are irrelevant, except some of
those with $l \geq 1, N=1, M=0,1$,
which are bounded by $C_{L,a} \epsilon \|g_n\|$.
Hence, we get (recall that $\|g\|\leq C\|f\|$ by (28))
\be
\|R_{L^n,K}g\|\leq C_{L,a} L^{-(1-\delta)n}
(\|f\|+\epsilon)
\en
which allows us to conclude the proof as before.
\hfill $\Box$

\vspace{3mm}

\no{\bf Proof of the Proposition}. To solve (10), we set
\be
h=A\psi
\en
and solve
\be
h=\p(a^*(\phi_0^*+A^{-1}h)\p(\phi^*_0+A^{-1}h))
\equiv \psi_0 + N(h)
\en
in the Banach space (13). We need some information
on the operator
$A$. In fact, $A$ is just the harmonic oscillator
in disguise:
\be
e^{{x^2\over 8}}Ae^{-{x^2\over 8}} = -\p^2+{x^2\over 16}+
{1\over 4} .
\en
Hence, the kernel of $A^{-1}$ is readily computed
from Mehler's
formula \cite{Si}. The result is
\be
A^{-1}(x,y) \equiv M(x,y)=\int_0^\infty dt M_t(x,y)
\en
with
\be
M_t(x,y)={ e^{-{t\over 2}}\over \pi^{\ha}(1-e^{-t})^{\ha}}
e^{-{1\over 4}{(x-e^{-{t\over 2}}y)^2\over 1-e^{-t}}} .
\en
In $N(h)$ we encounter terms $\p A^{-1}h$ and $\p^2 A^{-1}h$ .
We write the latter as $\p^2 A^{-1}h=(-A-\ha x\p) A^{-1}h
=-h-\ha x\p A^{-1}h$. Thus,
\be
N(h)=R(A^{-1}h,\p A^{-1}h,x\p A^{-1}h)
\en
with $R:{\bf C}^3\rightarrow {\bf C}$ analytic
near zero. Thus
all we now need is

\vspace{2mm}

\no{\bf Lemma.} {\it The operators $A^{-1}$, $\p A^{-1}$
and $x\p A^{-1}$ are continous in the norm {\rm (13)}.}

\vspace{2mm}
\no{\bf Proof.} We want to show
\be
|(1+|x|)\p^i(A^{-1}h)(x)|\leq Ce^{-{x^2\over 8}}\|h\|_N\; ,\;
i\leq N+1 .
\en
Let
\be
M=\int_0^\tau M_tdt+\int_\tau^\infty M_tdt\equiv
M^{(1)}+M^{(2)}
\en
For $M^{(2)}$ we have
\be
|\p^i(M^{(2)}h)(x)|\leq\int_\tau^\infty dt\int dy|
\p^iM_t (x,y)|
e^{-{y^2\over 8}}\|h\|_N
\en
 From (67) we have the estimate, for any $\delta >0$,
\be
|\p^iM_t (x,y)|\leq C\delta^{-\ha i}
e^{-{t\over 2}}(1-e^{-t})^{-\ha(i+1)}
e^{-{1-\delta\over 4}{(x-e^{-{t\over 2}}y)^2\over 1-e^{-t}}}
\en
Thus, using
\be
\int e^{-{1-\delta\over 4}{(x-e^{-{t\over 2}}y)^2\over 1-e^{-t}}
-{y^2\over 8}}dy=2\sqrt{\pi}((1-\delta){e^{-t}\over 1-e^{-t}}+\ha)^
{-\ha} e^{-{1-\delta\over 4}{x^2\over 1+(1-2\delta)e^{-t}}}
\en
we see that for $\tau$ large enough and $\delta$ small,
\be
|(1 +|x|)\p^i(M^{(2)}h)(x)|\leq C\|h\|_Ne^{-{x^2\over 8}}
\en

For $M^{(1)}$ we have (let $i=N+1$, the worst case),
integrating by parts,
\be
|\p^{N+1}(M^{(1)}h)(x)|=|\int^\tau_0 dte^{N{t\over 2}}
\int dy\p_x M_t (x,y)\p^N h|
\en
and thus by (72) and (73) this is bounded by
\be
|\p^{N+1}(M^{(1)}h)(x)|\leq {C\over\sqrt{\delta}}
\|h\|_N\int_0^\tau
dt \frac{e^{-{t\over 2}}}{(1-e^{-t})^{1/2}}
e^{-{1-\delta\over 4}{x^2\over 1+(1-2\delta)e^{-t}}}  f(t)
\en
where $f(t) = (\frac{1}{2} (1-e^{-t}) + (1-\delta)
e^{-t})^{-1/2}$ is ${\cal O} (1)$
for all $t$. Now, using $e^{-t}\leq (1+t)^{-1}$, we have
\be
-{1-\delta\over 4}{x^2\over 1+(1-2\delta)e^{-t}} \leq
-{x^2\over 8}-t{x^2\over 8}(\frac{1-2\delta}{2-2\delta+t})
\en
and so we get from the $t$ integral
\be
\int_0^\tau dt \frac{ e^{-ctx^2 -{t\over 2}}}{(1-e^{-t})^{1/2}}
\leq C   |x|^{-1}
\en
yielding
\be
|(1+|x|) \p^{N+1}(M^{(1)}h)(x)|\leq C\|h\|_N e^{-{x^2\over 8}} .
\non
\en
For $j<N+1$, write
\be
|\p^{j}(M^{(1)}h)(x)|=|\int^\tau_0 dt e^{j{t\over 2}}
\int dy M_t (x,y)\p^j h|
\en
and get, from (72) for $i=0$ and from (78),
\be
|(1+|x|)\p^{j}(M^{(1)}h)(x)|\leq  C\|h\|_N e^{-{x^{2} \over 8}} .
\en
The Lemma is proved.

\vspace{2mm}

Solving (64) is now trivial. We have
\be
\|\psi_0 \|_N \leq C_a \epsilon^{d+1}
\en
and
\be
\| N (h) \|_N \leq C_a (\epsilon^d \| h\|_N + \| h \|_N^2)
\non
\en

By the Lemma then
\be
\|\psi\|_N\leq C\|h\|_N
\non
\en
and the Proposition is proved.\hfill $\Box$

\vspace{5mm}

\no{\bf The perturbed Burgers' equation}

\vs{3mm}

Let us now discuss the perturbed Burgers' equation,
(3.36). To understand the relevant fixed point, consider
first the case $H=0$. Then, by the Cole-Hopf transformation
\cite{Bu}, the
Burgers' equation reduces to the heat equation : set
\be
\psi (x,t) = e^{\int^x_{-\infty} u (y,t) dy}
\en
Then, $\dot{\psi} = \psi''$ if
\be
\dot u = u'' + (u^2)'
\en
Now, $\lim_{x \rightarrow - \infty} \psi (x,1) = 1$,
 $\lim_{x \rightarrow
+ \infty} \psi (x,1) =  e^{\int^{+ \infty}_{-\infty} u
(y,1) dy}$, so that we look for a fixed point for $\psi$ of the form
(3). Using the fact that the inverse transformation
of (82) is $u(x,t) =
\frac{\psi'(x,t)}{\psi (x,t)}$, we get the family of fixed points
\be
f^*_A(x) = \frac{A e' (x)}{1 + A e (x)} = \frac{d}{dx}
\;\; \log (1 + A e (x))
\en
for the transformation (2.4) with $\alpha = 1$, where $u$
solves (83).

Note that the derivative in (84) explains why $f^*_A$ is
a fixed point of (2.4) with $\alpha = 1$ while
$\phi^*_0$ in (3) is a fixed point of (4) (where
$\alpha = 0$).

Since $e (- \infty) = 0, e (+ \infty) = 1$,
\be
\log (1 + A) = \int^{+\infty}_{- \infty} f^*_A (x) \; dx
\en
Now we can state

\vs{3mm}

\no{\bf Theorem 4} {\em Let $H : {\bf C}^3 \rightarrow
{\bf C}$ be analytic in a neighbourghood of $0$ with $d_F >
0$ and fix a $\delta > 0$. Then, there exists an
$\epsilon > 0$ such that, if $\| f \| \leq \epsilon$,
equation $(3.36)$ with $u(x,1) = f(x)$ has a unique
solution which satisfies
\be
\lim_{t \rightarrow \infty} t^{1 - \delta} \| u (\cdot
t^{1/2} , t) - t^{- 1/2} f^* _ A (\cdot) \| = 0
\non
\en
(the norms here are defined by $(3.3)$).}

\vs{3mm}

\no{\bf Proof}.
The proof is very similar to the one of Theorem 1, except
for the change of fixed point. Let
\be
\; \;\;\;\; f (x) = f^*_{A_0} (x) + g_0 (x)
\en
with
\be
\log (1 + A_0) = \int^{+ \infty}_{-
\infty} f (x) \; dx
\en
so that, by (85),
\be
\int^{+ \infty}_{- \infty} g_0 (x) \; dx =
 \hat g_0 (0) = 0
\en

We have $\| g_0 \| \leq C \| f \|$:
using $e' (x) = \frac{e^{- x^{2/4}}}{\sqrt{4 \pi}}$, we see
that $f^*_{A_0}$ and all its derivatives are integrable; using
$\| \hat h \|_\infty \leq \| h \|_1$ we have
\be
\| f^*_{A_0} \| \leq C |A_0 |
\en
Also,
\be
|A_0 | \leq C \| f \|
\en
for $\| f \| \leq
\epsilon$, because, by (87), $|A_0 | \leq C \| f \|_1
\leq C \| f \|$. For the last inequality, we use $\| f
\|_1 \leq C \| (1 + |x|) f \|_2$, by Schwartz' inequality,
 and  $\| f
\|_2 + \|x f \|_2 \leq  C \| f \|$, by Plancherel.

The local existence is proven as before.
To study $R_L$, let us write the solution $u(x,t)$ as
\be
u(x,t) = u^*_{A_0} (x,t) + v (x,t)
\non
\en
with
\be
u^*_{A_0} (x,t) = t^{- 1/2}  f^*_{A_0} (\frac{x}{\sqrt
t})
\non
\en
Since $u^*_{A_0} (x,t)$ solves (83) with $u^*_{A_0} (x,1)
=f^*_{A_0} (x)$, $v (x,t)$ satisfies :
\be
\dot v = v'' + ((u^*_{A_0} + v)^2 - (u^*_{A_0})^2)' +
H (u,u',u'')
\en
with
\be
v (x,1) = g_0 (x)
\en
Now,
\be
R_L \; f(\cdot) = L u (L \cdot , L^2)
= f^*_{A_0}(\cdot) + L \; v(L
\cdot , L^2) .
\non
\en
Write it as
\be
R_L \; f(\cdot) = f^*_{A_1}(\cdot) + g_1(\cdot)
\en
with $\int^{+ \infty}_{- \infty} g_1 (x) \; dx = 0$.

This means that we define $A_1$ (see (85)) by
\be
\log (\frac{1+A_1}{1+A_0}) = \int^{+ \infty}_{- \infty}
\; v (x, L^2) \; dx
\en
But, by (88), (92), $\int v (x,1) \; dx = 0$ and,
by (91),  $|\frac{d}{dt} \int v
(x,t) dx| \leq \int | H(u,u',u'')| dx \leq \|H(u,u',u'')\|$
which, by a bound similar to (3.14),(3.17), gives
\be
| \int^{+ \infty}_{- \infty} v (x,L^2) dx | \leq
C_{L,H} \| f \|^2
\en

So, inserting (95) in  (94), we have, for $\|f\|$ small,
\be
| A_1 - A_0| \leq C_{L,H} \| f \|^2
\en

We also have
\be
\| g_1 \| \leq C L^{- 1} \| g_0 \|+ C_{L,H}
\| f \|^2  \leq L^{-(1-\delta)}
\| f \|
\en
This is like (3.25) : writing (91) as an integral
equation we have $L v (L \cdot , L^2) = R_0 g_0 +$ rest;
$R_0 g_0$ contracts because of (3.4)
and the rest is the sum of two terms which are bounded
using (3.13): the first, coming from $2(u^*_{A_0}v)'$,
 is less than  $C \epsilon \| g_0 \|
\leq L^{-1} \|g_0\|$,
 (because, using (90),
$\|u^*_{A_0} \|_L +\|(u^*_{A_0})' \|_L \leq C \| f \| \leq C
\epsilon$), and the second (coming from $(v^2)' + H$) is
bounded by $C_{L,H} \| f \|^2$.

Now the iteration is exactly as before. The $H$ term runs down
with $L^{- n d_H}$ so that, using (95), we have
\be
| A_{n+1} - A_n | \leq C_{L,H} L^{- n d_H} \| f \|^2
\non
\en
and, as in (97),
\be
\| g_n \| \leq C L^{-(1-\delta)n} \| f \| .
\non
\en

The rest of the proof is as in Theorem 1.\hfill $\Box$

\vs{1cm}

\no{ \Large\bf 5. Discussion}
\addtocounter{equation}{-97}
\vs{1cm}

Now, we want to put our results in a more general framework,
relate them to other works, and
explain the analogy with the theory of critical phenomena.
The discussion will be heuristic or based on previously known results,
and we shall limit ourselves to the family of heat equations with
absorption:
\be
\dot{u}=u'' - u^{p}
\en
where $p>1$ is not necessarily an integer. In our terminology, $p>3$
is irrelevant, $p=3$ marginal, and $p<3$ relevant; $3$ has to be
replaced by $1+{2\over N}$ in
$N$ dimensions.

To understand the $p<3$ case, observe that (1) is
invariant under the scaling
transformation (2.3) for $\alpha ={2\over {p-1}}$. Finding a fixed
point for that transformation
amounts to finding a solution of the form
$u(x,t) = t^{-{1\over p-1}}f^* ({x\over {\sqrt t}}) $, where
$f^*$ is a solution
of the ODE (replacing $x\over {\sqrt t}$ by $x$)  :
\be
f'' +\ha x f' +{f\over {p-1}}- f^p=0
\en
The following results
are known about the solutions of (2):

1) For any $1<p<3$, there
exists an everywhere positive
solution $f^*_1$ of (2), which has almost
Gaussian decay at infinity \cite{Br1,Ga}.

2) For any $p>1$, there exists a
solution $f^*_2$ which decays at infinity like
$|x|^{-{2\over{p-1}}}$ \cite{Ga,KP1}.
Note that $f^*_2$ is
integrable only for $p<3$.

Rather
detailed results are known on the basin of attraction of the
various fixed points; we state them
loosely, see the references for more precise statements e.g. on
the type of convergence; also, in each case, the decay in time
of $u(x,t)$ is $t^{-\alpha\over 2}$, as in (2.10),
 where the exponent $\alpha$ is related
to the fixed point as in (2.3, 2.4).

1) For $p\geq
3$, and initial data $u(x,1)$ non-negative and integrable,
the asymptotic behaviour of the solution is governed by
$f^*_0$, as in Theorems 1, 2, with logarithmic corrections
for $p=3$ \cite{Ga,GV}. This is a global result
($u(x,1)$ is not assumed to be small); our results are
perturbative, and restricted to integer $p$ (but a general
non-linearity $F$),
but do not require $u(x,1)$ to be pointwise positive and
hold (when $p>3$) also if one has $+u^p$ in
(1); in that case, smallness of $u(x,1)$ is necessary since
large initial data blow up in a finite
time \cite{F,CEE,L}.

2) For $p<3$ and $u(x,1)$ non-negative and having (suitable)
Gaussian decay, the asymptotic
behaviour is governed by the non-trivial fixed point $f^*_1$
\cite{Ga,KP2}.

3) If one starts with $u(x,1)$ non-integrable and decaying at
infinity like $|x|^{-\alpha}$, with $0<\alpha <1$,
 then, for $p>3$, the relevant
"Gaussian" fixed point is $f^*_{\alpha}$ where
${\hat f^*_{\alpha}}(k) =|k|^{\alpha-1}e^{-k^2}$, which,
for any $\alpha$, is
a fixed point of (2.4), where $u$ solves the heat equation.
This fixed point has the right $|x|^{-\alpha}$ decay at
infinity.
Now $u^p$ is relevant, marginal, or irrelevant according to
whether $p< 1+{2\over \alpha}$, $p= 1+{2\over \alpha}$, or
$p> 1+{2\over \alpha}$. One knows \cite{KP1,GV} that, for a
non-negative initial data, the solution converges to
$f^*_{\alpha}$ for $p> 1+{2\over \alpha}$. For $p= 1+{2\over \alpha}$,
it converges to $f^*_2$. For $p< 1+{2\over \alpha}$,
the solution converges to
a  solution constant in space, $(p-1)^{-{1\over p-1}}t^{-{1\over p-1}}$,
which solves
(1) without the diffusive term: $\dot{u}= -
u^{p}$, and which can be viewed as a
(somewhat degenerate) new fixed point.

To see the analogy with the theory of critical phenomena,
consider an Ising model or a $\phi^4$ theory, on an $N$
dimensional lattice, at the critical point.
 $N>4$ is like $p>3$ here: $\phi^4$ is irrelevant and the
behaviour at the critical
point is governed by the Gaussian fixed point. For $N=4$,
$\phi^4$ becomes marginal, and the Gaussian behaviour
is modified, like here in Theorem 2, by logarithmic corrections.
However, this is
 not true for every marginal perturbation. In $\phi ^4$ theory, like
here for $p=3$, this happens because the marginal
term becomes irrelevant when higher order terms are included:
$A_n$ and, therefore, $f_n$ go to zero which is the same thing as
having a coupling constant in front of the $u^3$ term going to
zero. In particular, this higher order irrelevancy depends,
like in $\phi^4$ theory,
on the sign of the perturbation. For $+u^3$ in (1),
the solution blows up \cite{CEE,L}.
In point 3) above, the marginal perturbation ($p= 1+{2\over \alpha}$),
leads to a non-trivial fixed point, $f^*_2$ instead of $f^*_{\alpha}$.
Also, in the Burgers' equation (3.36), the
marginal term remains marginal to all orders and the solution is
governed by a new fixed point, which is however easy to write
down (see the end of Sect.4).

For $N<4$, $\phi^4$ becomes relevant and one expects the critical
behaviour to be governed by a non-trivial fixed point, whose
existence is however much harder to establish than here.
Another analogy with field theory concerns the constant
$A$ in (7): this is like  a "renormalised" constant whose
corresponding "bare" value is $A_0 ={\hat f}(0)$. One of the problems
 encountered in proving (7), instead of just a bound on $u(x,t)$,
 is that $A$, unlike $A_0$, may depend in a complicated way on
$f$
and $F$, and is not known explicitely. This usually limits the power
of ordinary perturbation theory: one may try to expand $u(x,t)$ around
$A_0 t^{-\ha} f^*_0$ which has  the wrong constant and the perturbation
series may not converge.
One of the advantages of the RG
method is that it allows to "build up" $A$ through a convergent sequence
of approximations, as we did in the proof. Also, note that in the marginal
case, the renormalisation of $A$ drives it to zero and produces the
logarithmic
correction in (39).

Finally, one may also interpret in RG
language the results on the existence of singular or very
singular solutions of (1). A solution is
$singular$ if, as $t\rightarrow 0$,
$u(x,t)$ becomes concentrated on a point, and it is
$very$ $ singular$ if, moreover, $\int dx u(x,t)$ diverges when
$t\rightarrow 0$.
For the family of equations (1), one knows
that, for $1<p<3$, there exists both a singular and a very
singular solution \cite{Br1,Br2}; the singular solution
is the fundamental one while
$u(x,t) = t^{-{1\over{p-1}}}f_1^*
({x\over {\sqrt t}})$ is very singular,
since $f_1^*$ is integrable. Besides,
the fundamental solution behaves, for $t\rightarrow 0$,
like the corresponding solution of the heat
equation $(t^{-{1\over 2}}f_0^* ({x\over {\sqrt t}}))$.
On the other hand, for $p\geq 3$, there does
not exist any singular solution \cite{Br2}.

This can be understood in
RG language: start with some non-singular
data at $t=1$; a singular solution would be obtained by solving (1)
backwards in time and letting $t\rightarrow 0$. This amounts to
running backwards the RG flow; hence, the
stability of the fixed points is inverted: $f_0^* $ becomes
stable for $1<p<3$ and unstable for
$p\geq 3$. So, for $p<3$, one would expect the solution to be attracted
to $f_0^* $ when $t\rightarrow 0$, thus explaining the presence
of a fundamental solution. On the other hand, for $p\geq 3$, one could only
go,
when $t\rightarrow 0$, towards $f_2^* $, which, because of its power law
decay at infinity, cannot be a singular solution (it is not integrable for
any $t$).
This correspondence between $t\rightarrow 0$ and $t\rightarrow \infty$
is similar to the
correspondence between "ultraviolet" and
"infrared" behaviour of fixed points in field theory and in critical
phenomena.

\vspace*{8mm}

{\bf {Acknowledgments}}

We thank P. Collet for explaining to us the problems discussed in Sect. 4
and we thank him and G. Papanicolaou for useful discussions. J.B. benefited
from the hospitality of Rutgers University and A.K. from the one of the
University of Louvain. This work was supported in part by NSF Grant
DMS-8903041.

\end{document}